\documentstyle[prl,aps,epsf]{revtex}
\newcommand{\bms}{\mbox{\boldmath $\sigma$ \unboldmath}}
\draft
\begin{document}
\twocolumn[\hsize\textwidth\columnwidth\hsize\csname @twocolumnfalse\endcsname
\title{Parity nonconservation in heavy atoms: The radiative correction enhanced
 by the strong electric field of the nucleus}
\author{A. I. Milstein$^{(a)}$ and O. P. Sushkov$^{(b)}$}
\address{$^{(a)}$ Budker Institute of Nuclear Physics, 
630090 Novosibirsk, Russia.\\
$^{(b)}$ School of Physics, University of New South Wales, Sydney 2052, 
Australia}

\maketitle

\begin{abstract}
Parity nonconservation due to the nuclear  weak charge is considered.
We demonstrate that the radiative corrections to this effect due to the 
vacuum fluctuations of the characteristic size larger than the nuclear 
radius  $r_0$ and smaller than the electron Compton wave-length $\lambda_C$  
 are  enhanced because of the strong electric field of the nucleus.
The parameter that allows one to classify the corrections is the large
logarithm $\ln(\lambda_C/r_0)$. 
The vacuum polarization contribution is enhanced by the
second power of the logarithm.
Although the self-energy and the vertex corrections do not vanish,
they contain only the first power of the logarithm.
The value of the radiative correction is 0.4\% for Cs and 
0.9\% for Tl, Pb, and Bi.
We discuss also how the correction affects the interpretation of the
experimental data on parity nonconservation in atoms.

\end{abstract}

]
Atomic parity nonconservation (PNC) has now been measured
in bismuth \cite{Bi}, lead \cite{Pb}, thallium \cite{Tl}, and cesium 
\cite{Cs}. Analysis of the data provides an important test of the 
Standard Electroweak model and imposes constraints on new physics beyond 
the model, see Ref. \cite{RPP}.
The analysis is based on the atomic many-body calculations
for Tl, Pb, and Bi \cite{Dzuba1} and for Cs \cite{Dzuba2,Blundell}.
Both the experimental and the theoretical accuracy is best for Cs.
Therefore,  this atom provides the most important information on
the Standard model in the low energy sector.
 The analysis performed in Ref. \cite{Cs} has indicated a deviation
of the measured weak charge value from that predicted by the Standard
model by 2.5 standard deviations $\sigma$.

In the many-body calculations  \cite{Dzuba1,Dzuba2,Blundell} the Coulomb 
interaction between electrons was taken into account, while the magnetic 
interaction was neglected.
The contribution of the magnetic (Breit) electron-electron interaction
was calculated in the recent papers \cite{Der,Dzuba3}. It proved to be
much larger than a naive estimate, and it  shifted the theoretical 
prediction for PNC in Cs. As a result, the  deviation from the 
Standard model has been reduced.
The calculations \cite{Der,Dzuba3} have already been used to get new 
restrictions on possible
modifications of the Standard model, see, e. g., Ref. \cite{Ros}.
The reason for the enhancement of the Breit correction has been explained
in Ref. \cite{Sushkov}.
In the case of the Coulomb residual interaction the effect of the 
many-body polarization is maximum for the outer electronic subshell and 
quickly drops down inside the atom \cite{Dzuba1,Dzuba2,Blundell}.
The Breit interaction is more singular at small distances than the Coulomb 
one. Hence, the polarization is maximum for the lowest
subshell ($1s^2$) and quickly drops down towards the outer shells.
The estimate of the relative effect of the magnetic polarization
gives $Z\alpha^2$ instead of naive $\alpha^2$, where $Z$ is the nuclear 
charge and $\alpha$ is the fine structure constant.
To find the Breit correction there is no need to repeat the involved
many-body calculations performed in Refs. \cite{Dzuba1,Dzuba2,Blundell}.
Indeed, the Breit correction comes from small distances, 
$r\sim a_B/Z$ ($a_B$ is the Bohr radius), while all the Coulomb
polarization and correlation corrections come from large distances, 
$r\sim a_B$. Therefore, it is sufficient to calculate the relative Breit
correction to some  PNC mixing matrix element (say $6s_{1/2}-6p_{1/2}$ 
mixing in Cs) in the simplest Hartree-Fock or RPA approximation. 
The relative Breit correction to the PNC effect with account of all many-body
Coulomb polarization and correlation corrections is exactly the same.

The Breit correction to PNC is just a part of the effect. This part is related 
to the virtual excitations of the $1s^2$ subshell. 
Another contribution comes from the vacuum fluctuations, i.e. from the
radiative corrections. Attempts to estimate this effect were
made in Ref. \cite{Lynn} and gave very small values of the corrections.
It has been pointed out recently \cite{Sushkov} that the strong electric field 
of the nucleus enhances the radiative corrections, and they may be
comparable with the Breit correction.
Very recently this suggestion has been confirmed  by the
numerical calculation of the vacuum polarization correction in Cs \cite{W}.

In the present paper we consider PNC in heavy atoms and calculate
radiative corrections enhanced by the strong electric field of the nucleus.
We calculate analytically the leading term in the correction and we also
estimate other terms.
It turns out that in Cs, Tl, Pb, and BI the radiative correction compensates 
the Breit correction calculated in Refs. \cite{Der,Dzuba3}. Thus,
we return back to the result of the experimental data analysis made in Ref. 
\cite{Cs}: deviation from the Standard model is 2.2 - 2.3 $\sigma$.

In the Standard model it is accepted to normalize the Weinberg angle
at the W-boson mass $M_W$. Atomic experiments correspond to a
very low momentum transfer compared to $M_W$. The renormalization from $M_W$ 
to zero momentum transfer was performed in Refs.\cite{Mar1,Mar2}. 
This renormalization is reduced to the  logarithmically enhanced single loop
corrections $\propto \alpha/\pi$ that lead to what is 
 usually called the radiative correction to the
nuclear weak charge. Account of this very important correction gives the
nuclear weak charge $Q_{W}$ measured in on-mass-shell electron 
scattering at zero momentum transfer, where  $p^2=m^2$ and $q=0$. 
However, atomic PNC corresponds to a different situation.
The electron in the strong nuclear electric 
field is off the mass-shell,  $p^2 \sim 1/r_0^2 \gg m^2$ and, besides,
the typical momentum transfer is of the order of the inverse nuclear radius,
$q\sim 1/r_0$. In a precise calculation these effects must be taken into 
account.
It is convenient to use $Q_{W}$ calculated in Refs. \cite{Mar1,Mar2} as a
reference point. Then the renormalization procedure is
the same as that in quantum electrodynamics and the correction vanishes
on mass shell at zero momentum transfer. In this approach it is clear
that the correction we are talking about is somewhat similar to the radiative 
correction to the hyperfine constant in a heavy atom \cite{KK}.

The wave function of the external electron is of the form
\begin{equation}
\label{Dirac}
u({\bf r})=
\left(
\begin{array}{c}
F(r)\Omega\\
iG(r)\tilde{\Omega}
\end{array}
\right),
\end{equation}
where $\Omega$ and $\tilde{\Omega}=-({\bms}\cdot{\bf n})\Omega$
are spherical spinors\cite{BLP}.
At small distances $r \ll Z\alpha\lambda_C$, where $\lambda_C$ is 
the electron Compton wave-length, the electron mass is small 
compared to the nuclear Coulomb potential, and the radial wave functions 
obey the equations
\begin{eqnarray}
\label{fg}
&&{{d(rF)}\over{dr}}+{{\kappa}\over{r}}(rF)-{{Z\alpha}\over{r}}(rG)=0,\\
&&{{d(rG)}\over{dr}}-{{\kappa}\over{r}}(rG)+{{Z\alpha}\over{r}}(rF)=0.\nonumber
\end{eqnarray}
For PNC effect we need to consider only
$s_{1/2}$ ($\kappa=-1$) and $p_{1/2}$ ($\kappa=+1$) electron states.
Solution of Eqs. (\ref{fg}) reads
\begin{equation}
\label{fg1}
F= Ar^{\gamma-1},\ \ \ 
G=A{{Z\alpha}\over{\kappa-\gamma}}r^{\gamma-1},
\end{equation}
where $\gamma=\sqrt{1-Z^2\alpha^2}$ and $A$ is some constant dependent
on the wave function behavior at large distances ($r\sim a_B$) \cite{Kh}.
In the leading approximation the  PNC interaction 
\begin{figure}[h]
\vspace{5pt}
\hspace{-35pt}
\epsfxsize=8.cm
\centering\leavevmode\epsfbox{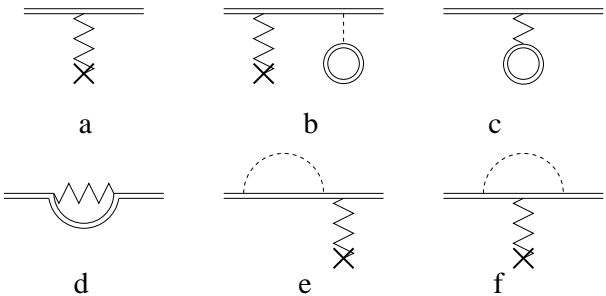}
\vspace{10pt}
\caption{\it {The leading contribution and the radiative corrections
to the PNC effect. The double line is the exact electron Green's
function in the Coulomb field of the nucleus, the cross denotes the nucleus,
the zigzag and the dashed lines denote Z-boson and photon, respectively.
}}
\label{Fig1}
\end{figure}
\noindent
related to the weak charge is due to  Z-bozon exchange, see  Fig.1a.
Calculation of the corresponding weak interaction matrix element 
gives \cite{Kh}
\begin{equation}
\label{pnc}
<p_{1/2}|H_{W}|s_{1/2}>_0=M_0\propto (F_sG_p-G_sF_p)|_{r=r_0}.
\end{equation}
At $r_0 \to 0$ this  matrix element is divergent, $M_0\propto r_0^{2\gamma-2}$.
As a result, the relativistic enhancement factor
is $R \approx$3 for Cs and $R \approx 9$ for Tl, Pb, and Bi \cite{Kh}. 
In the present paper we show that this divergence results in the double 
logarithmic enhancement of the radiative corrections.

The first correction is shown in Fig.1b. It corresponds to a modification
of the electron wave function because of the vacuum polarization.
In the leading $Z\alpha$  approximation the vacuum polarization
results in the Uehling potential \cite{Ueh}. At $r \ll \lambda_C$,
this potential is of the form
$V(r)\approx 2Z\alpha^2[\ln(r/\lambda_C)+C+5/6]/(3\pi r)$, 
where $C\approx 0.577$ is the Euler constant.
Account of higher in $Z\alpha$ corrections in the vacuum polarization
leads to a modification of the constant: 
$C \to C + 0.092Z^2\alpha^2+...$, see Ref. \cite{Mil}. However, this 
correction is small and can be neglected even for $Z\alpha \sim 1$.
The potential $V(r)$ modifies the Coulomb interaction in Eqs. (\ref{fg})
$-Z\alpha/r \to -Z\alpha/r + V(r)$. It is convenient to search for solution
of the modified Eqs. (\ref{fg}) in the following form ${\cal F}=F(1+F^{(1)})$, 
${\cal G}=G(1+F^{(1)})$, where $F$ and $G$ are given by (\ref{fg1}).
The functions $F_{s,p}^{(1)}$ and $G_{s,p}^{(1)}$ satisfy the
following equations
\begin{eqnarray}
\label{fg2}
&& \frac{1}{\gamma+\kappa}\,\frac{d F^{(1)}}{d x}- F^{(1)}+
 G^{(1)}=-\frac{2\alpha}{3\pi}\, x \nonumber \\
&&\frac{1}{\gamma-\kappa}\frac{d G^{(1)}}{d x}-G^{(1)}+ F^{(1)}
 =-\frac{2\alpha}{3\pi}\, x   \quad ,
\end{eqnarray}
where $x=\ln(\lambda/r)$, and $\lambda=\lambda_C\exp(-C-5/6)$.  The solution 
of these equations reads 
\begin{eqnarray}
\label{final}
&&  F^{(1)}=\frac{\alpha}{3\pi}\left[\frac{(Z\alpha)^2}{\gamma}x^2+
\frac{\kappa(\gamma+\kappa)}{\gamma^2}x+\frac{\kappa}{2\gamma^2}+a
\right]\nonumber\\ 
 &&  G^{(1)}=\frac{\alpha}{3\pi}\left[\frac{(Z\alpha)^2}{\gamma}x^2-
\frac{\kappa(\gamma-\kappa)}{\gamma^2}x-\frac{\kappa}{2\gamma^2}+a\right],
\end{eqnarray}
where $a$ is some constant.
Using ${\cal F}$ and ${\cal G}$ instead of $F$ and $G$ in Eq. (\ref{pnc})
we obtain the PNC matrix element in the form 
\begin{equation}
\label{pnc1}
<p_{1/2}|H_{W}|s_{1/2}>=M_0\left(1+\delta\right),
\end{equation}
where $\delta$ due to the diagram Fig.1b is
\begin{equation}
\label{ffgg}
\delta_b={{1+\gamma}\over{2}}\left(F_s^{(1)}+G_p^{(1)}\right)+
{{1-\gamma}\over{2}}\left(F_p^{(1)}+G_s^{(1)}\right).
\end{equation}
To find the  correction $\delta_b$ with the logarithmic accuracy
there is no need to calculate $a$ in Eqs. (\ref{final}), it is
enough to substitute the logarithmic terms from (\ref{final}) into
(\ref{ffgg}).  An analysis that also includes
 a consideration of distances $r\sim \lambda_C$ gives
\begin{equation}
\label{db}
\delta_b=\alpha\left({1\over{4}}Z\alpha+ {{2(Z\alpha)^2}\over{3\pi\gamma}}
\left[\ln^2(b\lambda_C/r_0)+B\right]\right),
\end{equation}
where $b=\exp(1/(2\gamma)-C-5/6)$, and $B\sim 1$ is some 
smooth function of $Z\alpha$ independent of $r_0$.
A numerical calculation of $\delta_b$ for Cs was performed recently in 
Ref. \cite{W}. The result is in a good agreement with Eq. (\ref{db}).
Comparison of Eq. (\ref{db}) with the result of Ref. \cite{W} allows also 
to determine $B$: $B\approx 1$.
We would like to emphasize that Eq. (\ref{db}) does not assume that 
$Z\alpha \ll 1$, it is valid for any $Z\alpha <1$. 
Note that the $(Z\alpha)^2$ term in (\ref{db}) is larger
than the  $Z\alpha$ one at $Z> 10$.

We already pointed out that the weak charge calculated in Refs.\cite{Mar1,Mar2}
corresponds to zero momentum transfer. 
On the other hand, it is clear from Eq. (\ref{pnc}) that the weak
interaction matrix element is determined by the momentum transfer
$q\sim 1/r_0$. The renormalization of the weak charge from $q=0$
to $q\sim 1/r_0$ is described by diagrams c and d in Fig.1.
A simple calculation gives the following correction
$\delta Q_W/Q_W=\delta_{cd}$ related to this renormalization 
\begin{equation}
\label{cd}
\delta_{cd}={{4\alpha Z}\over{3\pi Q_W}}
(1-4\sin^2\theta_W)\ln(\lambda_C/r_0)\approx -0.1\%.
\end{equation}
Where $\theta_W\approx $ is the Weinberg angle, $\sin^2\theta_W\approx
0.2230$, see Ref. \cite{RPP}.
Note that this correction is practically independent of $Z$
because $Z/Q_W \approx -Z/N\approx -0.7$, where $N$ is the number of neutrons.
One can also obtain the correction (\ref{cd})
using Eqs. (2a,b), and (3b) from Ref.\cite{Mar1}.

Next we consider the contribution of the electron self-energy operator 
$\Sigma$.
This operator being substituted to the Dirac equation, $m \to m+\Sigma$,
leads to the Lamb shift of the energy level and to the modification of
the electron wave function, see, e.g., Ref. \cite{BLP}. 
As shown in Fig.1e, this modification influences the matrix element of 
the weak interaction. The diagram Fig.1e is not invariant with respect to the
gauge transformations of the electromagnetic field.
However, the sum of the diagrams Fig.1e and Fig.1f  (the vertex correction) is 
gauge invariant. 
It is convenient to represent the self energy operator as a series
in powers of the Coulomb field of the nucleus,
$\Sigma=\Sigma_0+\Sigma_1+\Sigma_2+...$, see Fig.2. 
\begin{figure}[h]
\vspace{5pt}
\hspace{-35pt}
\epsfxsize=8.cm
\centering\leavevmode\epsfbox{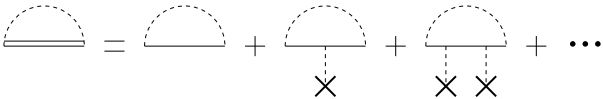}
\vspace{10pt}
\caption{\it {The electron self energy expanded in powers of the
Coulomb field. The solid line is the free electron Green's
function, the cross denotes the nucleus, and the dashed line
denotes the photon.
}}
\label{Fig2}
\end{figure}
\noindent
We need $\Sigma({\bf r},{\bf r'}|\epsilon)$ at $r \sim r' \ll \lambda_C$,
$\epsilon\approx m$. In this limit a calculation in the Feynman gauge
with logarithmic accuracy gives
\begin{equation}
\label{ss}
\Sigma_0={\hat p}{{\alpha}\over{4\pi}}\ln(p^2/m^2), \ \ \
\Sigma_1={{Z\alpha^2}\over{4\pi r}}\gamma_0\ln(p^2/m^2),
\end{equation}
where ${\hat p}=p^{\mu}\gamma_{\mu}$,
$\gamma_{\mu}$ is the Dirac matrix, and $p^{\mu}$ is the momentum operator.
All the higher terms are not logarithmically enhanced.
Further calculation in Feynman gauge is rather involved because the diagram
Fig.1f is also logarithmically enhanced and there is a delicate cancellation
between a part of the $\Sigma$-contribution and the logarithmic part of
Fig.1f. To avoid all these complications it is convenient to use the
Landau gauge. In this gauge neither $\Sigma_n$ ($n=0,1,2...$) nor the
contribution of Fig.1f contain the large logarithm $\ln(\lambda_C/r_0)$.
Therefore, after the renormalization, taking zero momentum transfer as
a reference point, the contribution of Fig.1f is of the form  
$\delta_f \sim Z\alpha^2(1+fZ\alpha/\pi+...)$,
where we expect $f\sim 1$.
Since $\Sigma $ does not contain logarithms,  at $r \ll \lambda_C$ we have
$\Sigma u =({{Z\alpha^2}/{\pi r}}){\cal D} u$, where
${\cal D} \sim1 $ is some matrix dependent on $Z\alpha$, and $u$ is the
electron wave function. Substitution of $\Sigma u$ into the Dirac
equation results in the $x$-independent terms of the order of
$\sim \alpha/\pi$ in the 
right hand sides of Eqs. (\ref{fg2}). As a result, the relative
correction to the matrix element of the weak interaction due to the
diagram Fig.1e contains logarithmically enhanced $Z^2\alpha^3$ terms.
There is also a $Z\alpha^2$ contribution coming from the distances
$r\sim \lambda_C$. All in all the total contribution of diagrams  
Fig.1e and  Fig.1f is of the form
\begin{equation}
\label{ef}
\delta_{ef} =Z\alpha^2\left[a_1+ a_2{{Z\alpha}\over{\pi}}
\ln(\lambda_C/r_0)\right].
\end{equation}
Our preliminary estimate gives $a_1 \approx 0.15$, and we expect
$a_2 \sim 1$. Therefore, the value of $\delta_{ef}$ for Cs is
$\delta_{ef}\sim 0.1\%$.
The calculation of the coefficient $a_2$ in Eq. (\ref{ef}) is a very 
interesting and challenging problem. 
At this stage we can claim only that the contribution (\ref{ef}) 
is much smaller than that of the Uehling potential (\ref{db}) because
it does not contain the big logarithm squared.

There is one more radiative correction that
has never been considered before. This contribution is due to the
virtual excitation of the nuclear giant dipole resonance $A^*$
shown in Fig.3.
\begin{figure}[h]
\vspace{-3pt}
\hspace{-35pt}
\epsfxsize=3.cm
\centering\leavevmode\epsfbox{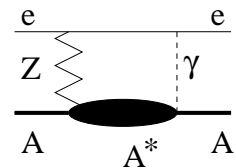}
\vspace{0pt}
\caption{\it {Virtual excitation of the giant nuclear resonance.}}
\label{Fig3}
\end{figure}
\noindent
Our estimate gives
\begin{equation}
\label{giant}
\delta_{A^*}\approx -0.1Z^{2/3}\alpha^2.
\end{equation}
So it is completely negligible.

Considering all the corrections, one has to remember about the contribution
of the electron-electron weak interaction. Although this is not a radiative 
correction,  we still denote it $\delta_{e-e}$. According to Ref. \cite{SF}
this correction is
\begin{equation}
\label{ee}
\delta_{e-e} \approx {{0.56}\over{ZR(Z)}}(1-4\sin^2\theta_W),
\end{equation}
where $R(Z)$ is the relativistic enhancement factor defined in Ref. \cite{Kh}.
Values of $R$ are presented after Eq. (\ref{pnc}). For Cs the 
$\delta_{e-e}$-correction is +0.04\%, so it is also negligible.

In summary, we have calculated  radiative corrections to the
parity nonconservation effects in atoms due to the strong electric
field of the nucleus. The structure of the correction is somewhat similar
to that for the radiative correction to the hyperfine interaction \cite{KK}.
However there are two essential differences. 1) The corrections to the
PNC effects are enhanced by the large logarithm dependent on the nuclear 
radius.
2) The contribution of the Uehling potential to the PNC radiative correction
is dominating because it contains the second power of the logarithm,
see Eq. (\ref{db}). The vertex and the self-energy corrections do not
vanish, but they are not enhanced by the second power of the logarithm.
The vertex, self-energy, nuclear excitations, and the electron-electron
corrections, see Eqs. (\ref{cd}), (\ref{ef}), (\ref{giant}), and (\ref{ee}),
contribute at the level $\pm 0.1\%$.

The correction (\ref{db}) has been derived for the electron-nucleus
weak interaction independent of the nuclear spin, i. e. for the weak
charge $Q_W$. However, the same formula is valid for the  PNC effect 
dependent on the nuclear spin, i. e. for the anapole moment.
This is obvious because the anapole moment interaction is
also local, see Ref. \cite{Kh}.

The radiative correction (\ref{db}) is $+0.4\%$ for Cs and
$+0.9\%$ for Tl, Pb and Bi. The value of the Breit correction in Cs is
$-0.6\%$, see Refs. \cite{Der,Dzuba3}. The Breit correction scales 
practically linearly with Z, see Ref.\cite{Sushkov}. Therefore, for Tl, Pb, 
and Bi the value of the Breit correction is $-0.9\%$. Thus, the radiative 
correction compensates the great part of the Breit correction in Cs, and 
practically exactly compensates it in Tl, Pb, and Bi.
With account of both corrections, the difference between the Standard
model prediction and the experimental result in Cs, see Ref. \cite{Cs},  
stays at the level of $2.2\sigma$.

Concluding we would like to say that both the radiative and the Breit 
corrections presently are known with high accuracy and this accuracy can 
be even further improved. 
The present limitation of the accuracy in the atomic
calculations comes from  many-body effects with pure Coulomb
electron-electron interaction considered in Refs. \cite{Dzuba2,Blundell}.
An improvement of the accuracy of these calculations is a feasible but
a very challenging problem. 

A.I.M. would like to thank M. G. Kozlov for helpful discussions, and the 
School of Physics at the University of New South Wales for warm hospitality 
and financial support (Gordon Godfrey fund) during a visit.
O.P.S. gratefully acknowledges discussions and correspondence with
W. R. Johnson and A. S. Yelkhovsky.


\begin{references}
\bibitem{Bi}  L. M. Barkov and M. S. Zolotorev, JETP Lett., {\bf 27}, 357 
(1978); M. J. D. MacPherson {\it et al.}, Phys. Rev. Lett., 
{\bf 67}, 2784 (1991).
\bibitem{Pb} D. M. Meekhof {\it et al.}, Phys. Rev. Lett., {\bf 71}, 3442 
(1993).
\bibitem{Tl} N. H. Edwards {\it et al.}, Phys. Rev. Lett., {\bf 74}, 2654 
(1995); P. A. Vetter {\it et al.}, Phys. Rev. Lett., {\bf 74}, 2658 (1995).
\bibitem{Cs} C. S. Wood {\it et al.}, Science {\bf 275}, 1759 (1997);
S. C. Bennett and C. E. Wieman, Phys. Rev. Lett.
{\bf 82}, 2484 (1999).
\bibitem{RPP} D. E. Groom {\it et al}
 Euro. Phys. J. C {\bf 15}, 1  (2000).
\bibitem{Dzuba1} V. A. Dzuba, V. V. Flambaum, P. G. Silvestrov, and
O. P. Sushkov, J. Phys. B {\bf 20}, 3297 (1987); Europhys. Lett.
{\bf 7}, 413 (1988).
\bibitem{Dzuba2} V. A. Dzuba, V. V. Flambaum, and O. P. Sushkov, Phys. 
Lett. {\bf 141A}, 147 (1989).
\bibitem{Blundell} S. A. Blundell, J. Sapirstein, and W. R. Johnson, Phys. 
Rev. Lett. {\bf 65}, 1411 (1990) and Phys. Rev. {\bf D45}, 1602 (1992).
\bibitem{Der} A. Derevianko, Phys. Rev. Lett. {\bf 85}, 1618 (2000).
\bibitem{Dzuba3} V. A. Dzuba, C. Harabati, W. R. Johnson, and M. S.
Safronova, Phys. Rev. A {\bf 63}, 044103 (2001);
M. G. Kozlov, S. G. Porsev, and I. I. Tupitsyn,
Phys. Rev. Lett. {\bf 86}, 3260 (2001).
\bibitem{Ros} J. L. Rosner, hep-ph/0109239 (2001).
\bibitem{Sushkov} O. P. Sushkov,  Phys. Rev. A {\bf 63}, 042504 (2001);
\bibitem{Lynn} B. W. Lynn and P. G. H. Sandars,
J. Phys. B {\bf 27}, 1469 (1994);
I. Bednyakov, L. Labzowsky, G. Plunien, and G. Soff,
Phys. Rev. A {\bf 61}, 012103 (1999).
\bibitem{W} W. R. Johnson, I. Bednyakov and G. Soff, private communication.
\bibitem{Mar1} W. J. Marciano and A. Sirlin, Phys. Rev. D {\bf 27}, 552 (1983).
\bibitem{Mar2} W. J. Marciano and J. L. Rosner,  Phys. Rev. Lett. {\bf 65}, 
2963 (1990).
\bibitem{KK} R. Karplus and A. Klein , Phys. Rev. {\bf 85}, 972 (1952);
see also K. Pachucki, Phys. Rev. A {\bf 54}, 1994 (1006).
\bibitem{BLP} V.B.Berestetskii, E.M.Lifshitz, and L.P.Pitaevskii,
{\it Relativistic quantum theory} (Pergamon Press, Oxford, 1982).
\bibitem{Kh} {\it Parity nonconservation in atomic phenomena},
I. B. Khriplovich, Gordon and Breach, 1991.
\bibitem{Ueh} E. A. Uehling,  Phys. Rev. {\bf 48}, 55 (1935).
\bibitem{Mil} A. I. Milstein and V. M. Strakhovenko, ZhETF {\bf 84}, 1247
(1983).
\bibitem{SF} O. P. Sushkov, V. V. Flambaum,
Yad. Fiz. {\bf 27}, 1307 (1978).
\end{references}
\end{document}